\documentclass[]{interact}

\usepackage{epstopdf}
\usepackage[caption=false]{subfig}

\usepackage[numbers,sort&compress]{natbib}
\bibpunct[, ]{[}{]}{,}{n}{,}{,}

\theoremstyle{plain}

\theoremstyle{definition}

\theoremstyle{remark}

\begin{document}


\title{A Start To End Machine Learning Approach To Maximize Scientific Throughput From The LCLS-II-HE}

\author{
\name{Aashwin Mishra, Matt Seaberg, Ryan Roussel, Fred Poitevin, Jana Thayer, Daniel Ratner, Auralee Edelen, Apurva Mehta}
\affil{SLAC National Accelerator Laboratory, Menlo Park, California, USA}
}

\maketitle

\begin{abstract}
With the increasing brightness of Light sources, including the Diffraction-Limited brightness upgrade of APS and the high-repetition-rate upgrade of LCLS, the proposed experiments therein are becoming increasingly complex. For instance, experiments at LCLS-II-HE will require the X-ray beam to be within a fraction of a micron in diameter, with pointing stability of a few nanoradians, at the end of a kilometer-long electron accelerator, a hundred-meter-long undulator section, and tens of meters long X-ray optics. This enhancement of brightness will increase the data production rate to rival the largest data generators in the world. Without real-time active feedback control and an optimized pipeline to transform measurements to scientific information and insights, researchers will drown in a deluge of mostly useless data, and fail to extract the highly sophisticated insights that the recent brightness upgrades promise.

In this article, we outline the strategy we are developing at SLAC to implement Machine Learning driven optimization, automation and real-time knowledge extraction from the electron-injector at the start of the electron accelerator, to the multidimensional X-ray optical systems, and till the experimental endstations and the high readout rate, multi-megapixel detectors at LCLS to deliver the design performance to the users. Applications with chained Machine Learning models represent a paradigm shift from conventional developments of digital twins for individual systems. For instance, errors and uncertainties from upstream models can severely affect the fidelity of downstream models and applications. Similarly, the sensitivity of any individual model in this chain can have deleterious ramifications on the overall performance. In this light, a holistic design approach where different stages can communicate is essential. Our approach is modular, but with emphasis on a co-design philosophy that aims to integrate every element impacting an experiment from injector-to-detector, from start-to-end. Our overarching philosophy is not to supplant, but to supplement human operators with data driven approaches for a domain-aware, risk-averse implementation. This start to end methodology focuses on common tools and techniques across the individual elements, such as software tools like Xopt, Badger, Lume, etc, along with uniform formats for data and model storage. Here, different modules learn and adapt based on best use cases from others. We test and train ML tools on common platforms for each element independently, then we chain these tools via real-time diagnostics, and finally integrate and optimize the entire chain. This is illustrated via examples from Accelerator, Optics and End User applications.
\end{abstract}


\section{Introduction}

The increasing brightness of light sources, including the diffraction-limited upgrade of the Advanced Photon Source (APS) and the high-repetition-rate upgrade of Linac Coherent Light Source (LCLS), opens pathways to get deeper understanding of the fundamental nature of the processes controlling the world around us, and opportunities to leverage those insights to gain unprecedented mastery of emerging new technologies and diseases plaguing human health.  However, these deep insights come with increasingly complex experiments requiring extreme precision over long durations. As an illustration, experiments at LCLS-II-HE will require the X-ray beam to be within a fraction of a micron in diameter, with pointing stability of a few nanoradians, at the end of a kilometer-long electron accelerator, a hundred-meter-long undulator section, and tens of meters of X-ray optics. Additionally, the increased complexity and throughput of experiments from enhanced brightness will produce data at over 1 Tb/s, rivaling the largest data generators in the world, requiring us to bridge the widening gap between the rate of data collection and that of scientific analyses. Without real-time active feedback control and an optimized pipeline to transform measurements into scientific information in nearly real-time, researchers will struggle to optimally utilize the high brightness sources, drown in a deluge of mostly useless data, and fail to extract the highly sophisticated insights that the upgrades promise.  Finally, the scientific success of light sources depends on large and varied user groups. There is an urgent need to render currently heroic experiments routine even for experienced light source users.   Additionally, there is a need to lower the barrier to success for domain experts with less experience at massive-scale data analysis or a nuanced knowledge of electrons and X-rays.  

Our vision, at the SLAC National Accelerator Laboratory (SLAC), is to develop AI-driven approaches to provide faster optimization and greater stability in a risk-averse manner for accelerating the discoveries from LCLS measurements. To achieve our vision, we aim to integrate components across the nearly 4 Km long LCLS X-ray laser, starting from the injector where high charge electron bunches are generated, the Linac where they are accelerated to several GeVs, the long undulator section where they are converted to attosecond to sub-picosecond high brightness X-ray laser pulses, to the X-ray optics which structures and guides the X-ray laser pulses to an experiment. We call our approach Start-2-End (S2E) , which will provide AI-driven experiment design and an analysis workflow to convert high-throughput measurements (billions of measurements a day) to accelerated delivery of discoveries in the material, chemical, and biological sciences.

Our overarching philosophy is not to supplant human operators, but to supplement them with data-driven approaches for a domain-aware, risk-averse implementation. Our philosophy also emphasizes retention of institutional knowledge, by letting the Machine Learning models learn formal “domain” knowledge, but also “gut-reactions” of  operators as well as the end users, in a human-in-the-loop approach. We aim to progress from operator augmentation, through trust-gaining exercises and learning from the intuition and experience of expert operators, by iterative progression to greater trust and more reliable automation. As machines become more risk-averse, we will incrementally deploy operator-tested and trusted automation and self-driving system modules, one at a time.  

\begin{figure}[h!t]
\centering
\includegraphics[trim={1.25cm 5cm 1.25cm 5cm},clip,width=\textwidth]{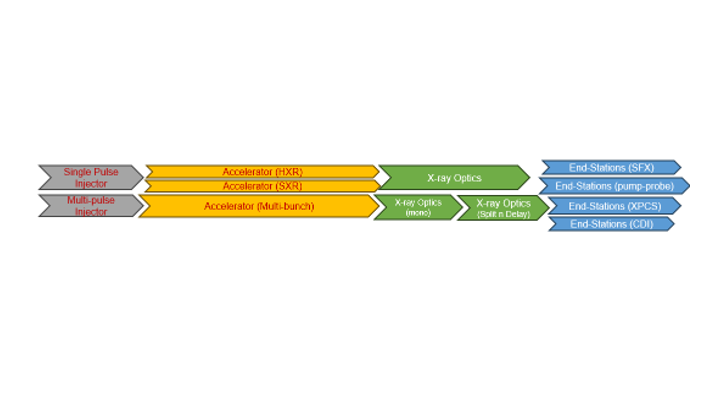}
\caption[]{Consolidated operational and model pipeline delineating each of the stages.}
\label{fig:fig1}
\end{figure}

We are developing the S2E framework by closely connecting local subsystems outlined in Figure \ref{fig:fig1}. Our approach is modular, but with emphasis on a co-design philosophy that aims to integrate every element impacting an experiment from the injector-to-detector, and to post-experimental analysis, from start-to-end. To achieve this we are building the S2E platform utilizing common AI and software engineering tools and techniques such as jointly developed software and control platforms like BlueSky \cite{bluesky}, Xopt \cite{xopt}, Badger \cite{badger}, Lume \cite{lume}, etc, along with uniform formats for data and model storage, and unified and prioritized access to high performance computing (HPC) resources. Here, different modules learn and adapt based on the best use cases from previously deployed modules. We test and train ML tools on common platforms for each element independently, then we chain these tools via real-time diagnostics, and finally integrate and optimize the entire chain. An S2E chain would often contain modules at different stages of operator control and autonomy, and gradually progress to higher levels of autonomy. All these have to be bolstered by requisite increments in computing infrastructure, along with concurrent software development efforts. 

To enable S2E modeling and control, machine learning models are initially trained on simulations, and then individually tested in a real system with operator supervision.  Once the operator assesses the robustness of a module, it is deployed in a stand alone manner.. These individual modules are then chained together to gradually create increasingly larger and more complex models, where the predictive outputs of models in earlier stages are the inputs for the models in the later stage, and outputs from later stages are included in feedback objectives to the AI model for earlier stages. Such chained data-driven models are commonplace in applications such as autonomous vehicles \cite{le2023comparing} and Multi-agentic workflows in Artificial Intelligence applications, but have also been utilized for Scientific analyses \cite{shafighfard2024chained, wahid2023multiphase, douglas2024uncertainty}. Conventional approaches, especially using end-to-end deep learning, consist of a single model accepting data, carrying out representation learning, and outputting the final predictions. Such models can be developed, tested and deployed in isolation. However, applications with chained Machine Learning models represent a paradigm shift from this conventional approach. For instance, any errors and uncertainties from upstream models can severely affect the fidelity of downstream models. In many cases, such errors can be magnified while propagating through downstream stages \cite{carreras2005introduction, attardi2015state}, referred to as Compounding or Cascading Errors. Similarly, the sensitivity of any individual model in this chain can have deleterious ramifications on the overall performance \cite{li2023error, gow2021uncertainty, eriksson2019scalable}, necessitating uniform testing of model stages during prototyping and integrated testing of the entire chain before deployment. 

In this light, a holistic design approach is essential, where different stages of modeling can communicate with and learn from each other, they can also be in close communication with the software engineers, and the data management team (including networking, as well as HPC experts), so the S2E deployed products function well together. 

Hereon, we outline and discuss details of this Start-to-End design and development approach, along with details of its constituent components. Then, we provide illustrative examples from different stages, including accelerators, focusing on accelerator optimization; complex photon optics, focusing on rapid alignment and enhanced stability; and end stations, focusing on inverse problems with tools for high data generation rate experiments.

\section{Start To End Machine Learning Approach}
Using the same platforms across the modules is essential for maintainability and debugging of the model pipeline, as well as for compatibility and integration for the modules constituting the pipeline. For instance, using the same platforms ensures that the same Application Programming Interfaces (APIs) are utilized in each module across all stages. This also renders the chain of models easier to maintain and debug, for instance, when updates are released for libraries. Additionally, this common framework ensures that the models from different stages are compatible, and the outputs of one model can be naturally used as inputs to the next. Based on this rationale, we are utilizing common platforms across stages and modules for the end-to-end modeling. These include libraries such as Xopt, Badger, Lume, etc. These libraries are developed at SLAC, thus enabling application-specific approaches and modifications to general techniques that are needed in the context of light source optimization and control.

Another cornerstone of this S2E approach is developing effective coordination between varying teams of subject matter experts modeling and controlling different physical phenomena inside the LCLS facility.. Due to differences in the physical laws that govern electron and X-ray beam dynamics the modeling and control of these beams is usually handled separately by different teams. However, in terms of modeling and control, problems in each of these subfields often share significant similarities \cite{gupta2021improving, edelen2019machine, mishra2023machine}. As a result, the same machine learning techniques and best practices can be used directly on or adapted to challenges in each. Adopting an S2E approach aims to unify research and development efforts in both the electron and X-ray modeling and control communities into a single package that has greater capabilities than the sum of its parts. These facets are highlighted as illustrative case studies in the next section.

The S2E approach breaks down the development, co-design and testing of the model chain in three phases. In the first phase, we develop and test the modules (models for individual stages) independently. This modular approach ensures that the models for different stages can be updated, for instance at a later date when better models are developed, when different settings for a stage are required, etc. This modular phase ensures that maintainability and testing of the entire chain is simplified, by allowing developers to test different modules in isolation. This incorporates human-in-the-loop systems, where an expert operator can “supervise” the model in initial stages. The operator can veto the model recommendations,  for instance if the recommendation may lead to a dangerous setting. This process is symbiotic, where the ML model can learn from the human feedback to improve its performance and learn the nuances of control of the specific system. This veto rate can be used as a metric to assess the trustworthiness of the model. Additionally, this interaction allows the operator to gain trust in the ML model. To this end, we are incorporating assistive visualizations to help the operator understand the rationale behind the ML model’s recommendations. This is particularly useful for complex algorithms such as optimization with learnt output constraints, for example for the ML model to learn to avoid sending the beam off of the screen during Bayesian Optimization based alignment.  

In the second phase, we chain these modules through real-time diagnostics.  An overarching need in chained models is to account for the predictive errors and uncertainties. Such Uncertainty Quantification (UQ) has been established as an essential component of Scientific Machine Learning (SciML) applications \cite{psaros2023uncertainty, koh2021evaluating, koh2023deep, mishra2021uncertainty}. However, most prior studies focus on epistemic uncertainties, that are due to model limitations, and aleatoric uncertainties, arising due to data limitations \cite{smith2024uncertainty, mishra2019uncertainty}. For chained models, in addition to these sources of uncertainty, we need to consider the compounding or cascading errors \cite{petersen2024uncertainty, petersen2024generalizing}, where the errors and uncertainties from upstream models can be magnified by the action of downstream models. Such uncertainty propagation studies can also identify specific models in the chain that are major contributors to the compounding   uncertainty, and effort can be focused on improving these models first. Additionally, uncertainty propagation studies ensure that prediction intervals for the overall model chain are estimated reliably. 

In the third and final phase, we integrate and optimize the performance of the entire model chain. In the longer term, this effort aims to move towards agentic frameworks that are incorporated in digital twins. Furthermore, this start to end approach can be utilized for Active Learning and Design of Experiments, where the model chain can guide future experiments with respect to the data that may be most beneficial to the meta-model. Additionally, the meta-model will be used to evaluate, rate and innovate new diagnostics. 

The S2E approach also encompasses the composition of the teams involved. At SLAC, in addition to Machine Learning Scientists and domain experts, the development teams include software engineers, as well as experts in High Performance Computing (HPC), Data transport, controls, etc. DevOps and MLOps workflows need to be designed, implemented and maintained for the success of this approach. This extends to partnerships for collaborative efforts as well. In addition to collaborations across different divisions at SLAC, we are partnering with experts at Stanford University, the Institute for Human-Centered AI, etc. These need to be augmented by collaborations across DoE laboratories as well as international collaborations. We are establishing these via programs like ML Exchange, ILLUMINE, etc. As an illustration, ILLUMINE is a SLAC-led effort consisting of different light and neutron sources, including Linac Coherent Light Source (LCLS)  and the Stanford Synchrotron Radiation Lightsource (SSRL) at SLAC, Advanced Light Source (ALS) at the Lawrence Berkeley National Laboratory, the Brookhaven National Laboratory, the Argonne National Laboratory (ANL) and the Oak Ridge National Laboratory (ORNL). This effort aims to build a modular framework to close the loop between fast analysis, ML-assisted decision making, and data acquisition to drive experiments on diverse timescales ranging from seconds to days. ML Exchange is a multi-facility initiative, collaborating with the Advanced Photon Source and Center for Nanoscale Materials (ANL), Linac Coherent Light Source (SLAC), National Synchrotron Light Source II (BNL), and Center for Nanophase Materials Sciences (ORNL) to integrate ML into experimental workflows, with a focus of Ptychography \cite{hoidn2023physics, vong2024portable} evinced in the Ptychodus software being developed in this effort.

\section{Illustrative Case Studies}

\subsection{Accelerator Application}
There are numerous optimization and control tasks associated with operating accelerators. At LCLS, the accelerator settings must be re-configured for each set of user experiments that require different photon beam characteristics. This involves changing the accelerator beam energy, the shape of the beam in time and energy (i.e. the “longitudinal phase space”), and in some cases conducting highly customized setups for advanced needs (e.g. attosecond pulses, multiple pulses, etc). Increasingly, users are interested in more dynamic control over beams, such as being able to adjust the time separation and energy separation of two beams on demand while controlling the shape of each. At present, operators often tune in a sequence along the accelerator: injector settings are adjusted to minimize emittance at the injector, the beam’s longitudinal phase space is adjusted by observing images on a transverse deflecting cavity and adjusting RF cavity phase and amplitude settings, and the FEL pulse intensity is adjusted through quadrupole focusing magnets. The final tuning to adjust the beam’s spectra and account for changes over time can often involve numerous adjustments to each of these sets of variables. Overall, there are hundreds of variables that are regularly adjusted by operators during tuning and thousands that are monitored.

\begin{figure}[h!t]
\centering
\includegraphics[trim={5cm 0 5cm 0},clip,width=1\textwidth]{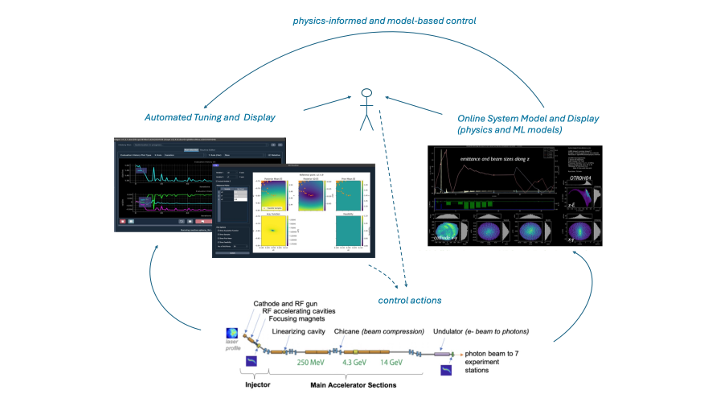}
\caption[]{Accelerator tuning tools and live system models (digital twins) are available for operators to use during setup of new configurations. }
\label{fig:fig2}
\end{figure}

We have been developing tools for aiding automated tuning in the accelerator control room, as well as for predicting/tracking machine behavior over time and enabling model-based control via the use of digital twins. Our ML-based tuning software, consisting of the Xopt algorithm driver and the Badger user interface, enable operators to choose from traditional and ML-based optimization and control algorithms and apply these to chosen combinations of variables, objectives, and constraints (including learned output constraints, such as learning to avoid sending a beam off of a screen). We are also combining physics and ML-based models into a digital twin platform that interfaces with our local HPC system, the S3DF, and NERSC. The LCLS-II injector online model, for example, was used to supplement physicists’ intuition by predicting machine behavior at points where we do not have diagnostic measurements; this enabled improved hand-tuning to find a good initial starting point for further fine-tuning with ML. This combination yielded the best emittance then seen during LCLS-II injector commissioning. 

The accelerator tuning problem itself can be considered as a template for the overall S2E approach: we are increasingly moving toward enabling accelerator tuning to occur over a broader range of variables and subsystems simultaneously, and helping inform these tuning approaches with comprehensive system models of the accelerator. We are now exploring using these same tools in conjunction with photon beamlines. Traditionally, accelerator and photon beamline tuning have been conducted separately, but we would like to link the two processes more directly. Start-to-end system models and shared tuning tools between the accelerator and photon beamline portions are needed and are under active development, and existing challenges with regard to linking photon and accelerator data are being addressed.

\subsection{Optics Application: Hard X-Ray Split And Delay Optimization}

\begin{figure}[h!t]
\centering
\includegraphics[width=\textwidth]{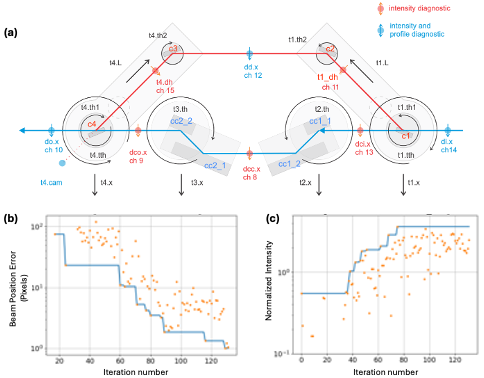}
\caption[]{Optimization of the HXRSND (a) Outlines the Split And Delay system with the CC branch in blue and the delay branch in red, (b) Beam Position Error minimization using TuRBO for the experimental beamtime, (c) Intensity maximization using TuRBO for the same run.}
\label{fig:fig3}
\end{figure}

While discussing the optimization and control of complex optics systems at SLAC, we use the Hard X-Ray Split and Delay (HXRSND) as an example. As outlined in Figure \ref{fig:fig3} (a), this consists of two branches: the minimally adjustable “channel-cut” (CC) branch outlined in blue, and the “delay” branch with twelve degrees of freedom outlined in red. This delay range lies between 5 to 500 ps. With the advent of LCLS-II-HE, this HXRSND optics will be critical for ultrafast studies of complex materials. We, therefore, must ensure that operational inefficiencies and system limitations do not become a bottleneck for these experiments. The alignment of the HXRSND requires a spatial overlap between the two branches at the sample with very high precision, along with optimized intensity at the output.  As an illustration, for X-ray Photon Correlation Spectroscopy (XPCS) where the HXRSND can provide twin pulses with custom delays, both the branches must be aligned to the same photon energy to within ~0.1 eV, with almost perfect overlap and matched intensities from the two branches. Currently, the HXRSND is aligned sequentially (from the right to the left in Figure \ref{fig:fig3} (a)), by an experienced system expert, using intermediate sensors, because even an experienced human is unable to optimize in more than 2-3 dimensions. It takes an experienced operator between 1 to 4 hours for alignment, and the final setting is often far from optimal. 

Multidimensional data-driven Bayesian Optimization is a promising approach for rapid  alignment of the HXRSND to the global optimal. In keeping with the S2E approach, we utilise a common platform for modeling and control, specifically Xopt. Additionally, we utilize best practices learnt from Accelerator optimization to guide our choice of optimization algorithm. 

The HXRSND optimization involves crystal optics, which have a singular optimal, bringing unique challenges. The optimal alignment of a crystal optics (in the scattering plane) is characterized by a sharp rise, followed by a gently sloping top (the Darwin curve).  Finding the optimum setting on top of these sharp, flat-topped regions requires submicron and nanoradian precision in multidimensional manifolds that are many orders of magnitude wider than the optimum along each dimension, becomes a “needle in a haystack” problem, and presents a significant challenge not only for humans but also for traditional Bayesian optimization (BO) approaches. We find that conventional Bayesian optimization, both single and multi-objective, was unable to converge to an acceptable optimum even after a few hundred iterations. 

In keeping with the needle in a haystack nature of the optimization, we need a strategy that is focused on not just finding an optimum, but also on gradually reducing the search manifold. Following experience with accelerator optimization, we adopted an algorithm known as Trust Region Bayesian Optimization (TuRBO) \cite{eriksson2019scalable}. TuRBO utilizes sets of local models with sample allocation across trust regions to refine the initial trust region. Using TuRBO with a single scalarized objective involving both the intensity and overlap, we are able to achieve consistent high-quality optima quickly in simulations as well as in real-world experimental conditions during beamtimes. Representative results from the beamtime are reported in Figure \ref{fig:fig3}, (b) and (c). The beam width for the experiment was 10 pixels. As can be seen in Figure \ref{fig:fig3} (b), TuRBO can consistently achieve a minimum of under 1 pixel inside 130 samples.  Similar results are observed for the intensity optimization, where TuRBO achieves an intensity maximum better than the optimum manual setting within 100 samples. (The orange points are all of the trials and the blue lines connect the best found optima.) TuRBO reduces the alignment time of the HXRSND system from the current 1-4 hours to the order of a few minutes, where most of the alignment time is spent in moving the motors.  It also aligns in all 12 dimensions simultaneously and does not require intermediate sensors.

\subsection{Endstation Application: Revolutionizing Structural Dynamics Research Via The autoMFX Initiative and Its Key Components}

The autoMFX initiative is set to transform structural dynamics studies at the Macromolecular Femtosecond Crystallography (MFX) instrument of the Linac Coherent Light Source (LCLS) \cite{sierra2019macromolecular, boutet2016new} by enhancing accessibility, automation, and real-time feedback in experimental processes. This comprehensive strategy aims to significantly increase the efficiency of experiments while broadening the user base, ultimately facilitating advanced biomedical research through a more streamlined workflow that connects accelerator controls to data analysis. At the heart of the initiative are three primary goals. The first involves the development of integrated automation frameworks. By leveraging existing Experimental Physics and Industrial Control Software (EPICS), autoMFX intends to automate requests for beam adjustments, thereby improving communication protocols between the Accelerator Directorate and X-ray instruments. This optimization will also extend to beamline alignment, enhancing the responsiveness of experimental setups. The second goal is to improve data processing workflows, which centers on establishing the LCLS Unified Task Executor (LUTE). This framework aims to automate complex data analyses across various X-ray instruments. LUTE will provide real-time feedback to confirm the efficacy of time-resolved studies, facilitating swift decision-making and consolidating both X-ray Emission Spectroscopy (XES) and Serial Femtosecond Crystallography (SFX) workflows. The consolidation is intended to promote seamless data integration, thereby enabling more efficient execution of experiments. The third goal focuses on enhancing user accessibility and experimental efficiency. To achieve this, the initiative seeks to simplify experimental setups through standardized protocols and user interfaces. By lowering the barrier for non-expert users, autoMFX will empower a broader range of researchers to engage in cutting-edge XFEL experiments with reduced specialized knowledge. The expectation is that automated protocols will yield improved data quality and enhanced reproducibility while accelerating the turnaround time for data analysis.

\begin{figure}[h!t]
\centering
\includegraphics[trim={4cm 0 4cm 0},clip,width=\textwidth]{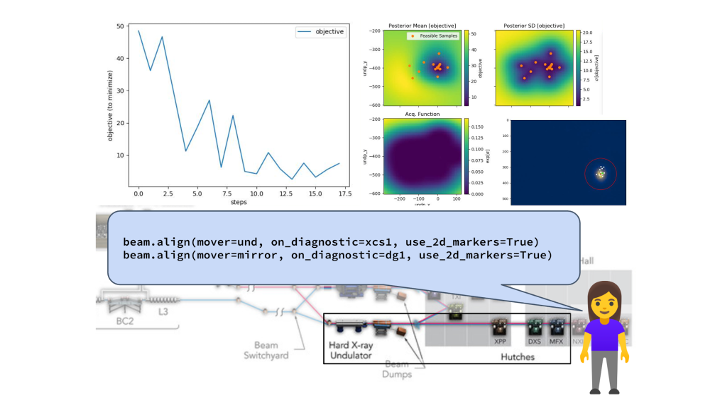}
\caption[]{Schematic outline of running Bayesian optimization based beamline alignment at MFX using Xopt.}
\label{fig:fig4}
\end{figure}

The autoMFX initiative outlines a series of milestones spread over the coming years. The development of a cohesive client-server model across the Accelerator and Photon Science Directorates at SLAC aimed at automating the sharing of diagnostic data is a primary focus. Additionally, routine optimization of X-ray beam parameters will be demonstrated, alongside the integration of hutch python, Bluesky, and Xopt to enhance operational frameworks. Ultimately, the initiative aims to achieve complete automation of LUTE workflows for SFX and XES, making these workflows available for use by multiple external research groups and establishing a reliable framework for ongoing operations. Central to autoMFX is the integration of advanced tools such as Xopt, AMI2, and LUTE, all of which will enhance operational efficiency across the LCLS facility. Xopt serves as an advanced machine-learning tool designed for optimizing beam parameters through Bayesian optimization techniques. By analyzing the relationships between optical adjustments and beam qualities, Xopt can facilitate rapid convergence to optimal configurations. Its integration within the Bluesky framework underpins much of the automation strategy, particularly in minimizing manual interventions and streamlining beamline adjustment processes. AMI2, or the Analysis and Monitoring Interface for LCLS-II, is a sophisticated framework designed to facilitate real-time data analysis and visualization for experiments conducted at the Linac Coherent Light Source (LCLS). It provides researchers with essential tools for monitoring experimental parameters and analyzing data streams, thereby enabling timely decision-making during complex scientific experiments. This framework consolidates data outputs from the Data Acquisition (DAQ) system into actionable insights, improving the conversion of data streams from EPICS into relevant feedback for experimental adjustments. As envisioned, LUTE will be the center for automating the intricate data analysis workflows utilized in the X-ray instrumentation at LCLS. Beyond managing routine data processing tasks with real-time analytics, LUTE is expected to evolve to eventually supplant the existing AMI2 framework for more resource-intensive analytical tasks.

The autoMFX initiative holds immense potential to democratize access to advanced research facilities at LCLS by reducing setup complexities and minimizing human error through automation. This transformation will nurture a new era of experimentation—from real-time adjustments in beam parameters to the facilitation of sophisticated data processing. Ultimately, the initiative is poised to significantly enhance the pace and accessibility of scientific discovery within the realm of biomedical research. In summary, the integration of automation frameworks, particularly through tools like Xopt, AMI2, and LUTE, signals a revolutionary progression in structural dynamics studies at LCLS. With this initiative, researchers are expected to achieve greater efficiency, accuracy, and accessibility in their investigations, ushering in an exciting future for the scientific community as it seeks to unravel the complexities of biological systems using cutting-edge XFEL technology.

\section{Conclusions}
The increasing brightness of light sources like APS and LCLS is enabling more complex experiments that require exceptional precision and stability, with X-ray beams focused to sub-micron diameters and nanoradian pointing stability. This advancement comes with challenges including unprecedented data generation rates that threaten to overwhelm researchers without proper systems for real-time feedback control and knowledge extraction, potentially preventing scientists from realizing the full benefits of these upgrades.

SLAC is addressing these challenges through a comprehensive S2E (Start-to-end) Machine Learning approach that is outlined and detailed in this article. This approach spans from the electron-injector through X-ray optical systems to experimental endstations and detectors. Their strategy employs chained Machine Learning models in a holistic design framework where different components communicate effectively, supplementing rather than replacing human operators. The approach uses common tools and standardized data formats, with modules that learn from each other's use cases. Implementation follows a progressive methodology: training tools independently for each component, chaining them via real-time diagnostics, and finally optimizing the entire integrated system from injector to detector. We exemplify the application of this approach at SLAC with illustrative examples ranging from accelerators, photon optics and endstations. 

\textbf{Acknowledgements}: Use of the Linac Coherent Light Source (LCLS), SLAC National Accelerator Laboratory, is supported by the U.S. Department of Energy, Office of Science, Office of Basic Energy Sciences under Contract No. DE-AC02-76SF00515.

\bibliographystyle{tfq}
\bibliography{interacttfqsample}

\end{document}